\documentclass[aps, prd, superscriptaddress, groupedaddress, showkeys, showpacs, nofootinbib]{revtex4-2}
\usepackage{epsfig}
\usepackage{amssymb}
\usepackage{amsmath}
\usepackage{amsfonts}
\usepackage{graphicx}
\usepackage{mathrsfs}
\usepackage[dvipsnames]{xcolor}
\usepackage{multirow}
\usepackage{lipsum}
\usepackage{booktabs}
\usepackage{braket}
\usepackage{hyperref}
\usepackage[toc]{appendix}
\usepackage{tikz}

\newcommand{\cL}{\mathcal{L}}

\newcommand{\be}{\begin{equation}}
\newcommand{\ee}{\end{equation}}
\newcommand{\bea}{\begin{eqnarray}}
\newcommand{\eea}{\end{eqnarray}}

\newcommand{\ed}{\end{document}}

\newcommand{\bi}{\begin{itemize}}
\newcommand{\ei}{\end{itemize}}

\newcommand{\bce}{\begin{center}}
\newcommand{\ece}{\end{center}}

\begin{document}

\title{Non-Hermitian Gravitational Wave Scattering}

\author{Burak Pekduran}\email{burak.pekduran.bp@gmail.com}
\affiliation{Department of Physics, Istanbul University, 34134, Vezneciler, Istanbul, Türkiye}
\author{Mustafa Sar{\i}saman}\email{mustafa.sarisaman@istanbul.edu.tr}
\affiliation{Department of Physics, Istanbul University, 34134, Vezneciler,
Istanbul, Türkiye}
\affiliation{National Intelligence Academy, Institute of Engineering and Science, Ankara, Türkiye}

\author{Ekrem Ayd{\i}ner}\email{ekrem.aydiner@istanbul.edu.tr}
\affiliation{Department of Physics, Istanbul University, 34134, Vezneciler,
Istanbul, Türkiye}

\affiliation{Department of Physics, Koç University, Sarıyer, 34450, İstanbul, Türkiye}

\begin{abstract}

In this study, the non-Hermitian scattering of gravitational waves is examined, and their behavior at spectral singularities is discussed. We investigate the non-Hermitian properties of gravitational waves through the construction of a transfer matrix. By examining spectral singularity points obtained from the transfer matrix, we explore the behavior of gravitational waves at these spectral singularity points and compare the theoretical results with observed measurements. Our findings demonstrate that the frequency values of gravitational waves measured on Earth at spectral singularity points exhibit perfect agreement with the corresponding Hubble constant values. This alignment underscores the significance of the non-Hermitian characteristics of gravitational waves. We anticipate that this research will contribute to a deeper understanding of the role of non-Hermitian phenomena in gravitational physics and provide a foundation for future studies in the field.
\medskip

\end{abstract}

\vspace{2mm}  

\vspace{2mm}

\maketitle

\section{Introduction}

Gravitational waves are defined as ripples in the curvature of space-time that propagate outward from gravitational sources at the speed of light \cite{gw1, gw2, gw3, gw4, gw5, gw6, gw7, gw8, gw9, gw10, gw11}. These waves, predicted in Albert Einstein's Theory of General Relativity\cite{gw1}, are formed especially by the effect of accelerating massive objects. For example, astronomical events such as binary star systems orbiting each other are among the powerful sources of gravitational waves \cite{gwb1, gwb2, gwb3}. The existence of these waves was directly detected by observations made by NSF LIGO \footnote{LIGO stands for the Laser Interferometer Gravitational-Wave Observatory in the United States.} in 2016\cite{Abbott_2016,ligo, ligo2, ligo3, ligo4}. These observations have also been confirmed by Virgo and KAGRA\cite{Abbott_2017,Virgo,Kagra}.

This historic journey began on September 14, 2015, when LIGO made its first physical measurement of the ripples in space-time generated by the gravitational waves produced by the collision of two black holes, named GW150914, located 1.3 billion light-years away. This groundbreaking detection by NSF LIGO marked one of the most significant scientific milestones in human history. The lead researchers behind this achievement, Rainer Weiss, Kip Thorne, and Barry Barish, were awarded the Nobel Prize in Physics in 2017 for their crucial contribution to the discovery of gravitational waves. The later LIGO and Virgo detection
\cite{Abbott_2017} of the Binary Neutron Star merger GW170817 and electromagnetic follow-ups were relevant to constraining cosmological models.

It is known that the universe is full of extraordinarily massive objects that produce gravitational waves. These include pairs of black holes or neutron stars orbiting one another, as well as systems consisting of a neutron star and a black hole, or even massive stars that end their lives by either collapsing or transforming into new forms. We now know that the type of collision determines the characteristics of the gravitational waves generated. These waves typically manifest in different forms: continuous, compact binary spirals, stochastic, and bursts. Moreover, the different types of collision suggest that they produce gravitational waves with distinct reaction times. For example, the first black hole merger detected by LIGO emitted a signal lasting only a fraction of a second. In contrast, the first neutron star merger detected by the LIGO and Virgo collaboration in August 2017 produced a signal that could be detected for more than 100 seconds. These time spans likely depend on factors such as the energy and distance of the colliding objects. Although black hole mergers release the highest amounts of energy and intensity in the universe, the gravitational waves reaching our detectors from millions or billions of light-years away are far weaker. The gravitational waves detected by LIGO caused space-time ripples that were 10,000 times smaller than the size of an atomic nucleus, highlighting the extraordinary sensitivity required for their detection.

Historically, researchers have primarily relied on electromagnetic radiation, such as visible light, X-rays, radio waves, and microwaves, to explore the universe. Recently, however, there has been growing interest in using neutrinos to gather information about the cosmos. In contrast to these methods, gravitational waves offer a new way to explore the depths of the universe. By tracking gravitational waves, we can examine the distant past of the universe, gain insights into its cosmic beginnings, and trace the aftermath of galactic collisions. More significantly, gravitational waves interact very weakly with matter similar to WIMP \cite{borah2022probing, lu2022probing, bringmann2024hunting, roszkowski2018wimp} —unlike electromagnetic radiation, which can be absorbed, reflected, refracted, or bent by gravitational forces. As a result, gravitational waves travel across the universe almost undisturbed, carrying undistorted information from their origins. By detecting and analyzing these waves, we will be able to observe the universe in ways that were previously unimaginable. For instance, gravitational waves enable the study of systems like binary black holes, which are invisible to electromagnetic radiation, and provide new insights into objects such as neutron stars that would otherwise be inaccessible.

The mathematical description of gravitational waves is made by expanding Einstein's field equations with a weak metric perturbation\cite{gwth1, gwth2, gwth3}. The Einstein field equation is a nonlinear equation with extraordinary mathematical and physical symmetries that describe how the energy and momentum of matter interact with the geometry of spacetime. The existence and propagation of gravitational waves arise as one of the special solutions of these equations and expresses perturbative changes in the curvature of spacetime\cite{gwth3}. The detection of these waves once again proves the accuracy of general relativity and opens a new window to the understanding of the universe.

Studies on gravitational waves are also essential to a broad range of scientific fields, from astrophysics to quantum physics, as these waves carry information about the universe’s most fundamental physical processes and provide a means to test many theories in modern physics \cite{gw2, gw4, gw5, gw7, gw9, gwth1, gwth2, gwth3}. 
Gravitational waves can play an important role beyond testing gravitational theories. For example, recent detections have been used to estimate an initial value for the Hubble parameter \cite{nature-2017a,Mukherjee_2021,arxiv-2020a,arxiv-2020b}.

In addition, gravitational wave scattering is another significant physical phenomenon. The scattering of gravitational waves by massive objects serves as a crucial tool for testing Einstein’s general theory of relativity, and has led to strong confirmation of the theory. Moreover, gravitational wave scattering provides valuable insights into the deeper, less understood aspects of gravitational waves and the nature of their sources. \cite{gws1, gws2, gws3, gws4, gws5, gws6, gws7, gws8}. Gravitational waves are generated by various sources, and their waveforms carry information about these origins. As a result, perturbation methods used in scattering analysis are expected to play a vital role.

In this study, we adopt a novel approach to investigate gravitational wave scattering. Specifically, we analyze these waves within the framework of recently developed non-Hermitian physics, employing scattering formalism. To this end, we construct a theoretical model of non-Hermitian gravitational wave scattering using the transfer matrix method. The real zeros of specific components of the transfer matrix reveal exceptional points that are characteristic of non-Hermitian physics. These points, in our case, referred to as spectral singularities, are known to generate purely outgoing waves, as established in the literature. Our configuration is given in~Fig.~\ref{fig1}, which consists of a representative gravitational wave source generating ripples around it in the radial direction.
  
On the other hand, it should be noticed that non-Hermitian physics investigates systems where energy and other observables exhibit non-conservative dynamics, often due to the presence of complex ingredients (like gain and/or loss). In such systems, traditional quantum mechanical concepts are modified, leading to phenomena such as exceptional points \cite{bender, ijgmmp-2010, longhi4, longhi3, nonhermit1, nonhermit2, nonhermit3, nonhermit4, nonhermit5, nonhermit6, nonhermit7, nonhermit8, nonhermit9, nonhermit10, nonhermit11, nonhermit12, nonhermit13}. At these exceptional points, although the system may possess real eigenvalues, the eigenstates can merge. Although non-Hermitian physics has gained significant attention in optics, solid-state physics, condensed matter systems and various other fields of physics in recent years \cite{biophys, biophys2}, it has largely been overlooked in the field of gravitational physics for reasons that remain unclear. In the context of scattering, the spectral singularities of optical systems correspond to states with divergent reflection and transmission amplitudes for real $k$-values, resulting in zero-width resonances and laser threshold states. These conditions typically generate purely outgoing waves \cite{p123, prl-2009, CPA, lastpaper, pra-2011a, pra-2012a}. This behavior is a natural consequence of non-Hermitian physics, and we expect similar behavior in gravitational waves. In recent years, significant progress has been made in exploring the new phenomena and insights enabled by non-Hermitian physics, with numerous fascinating studies currently underway \cite{nonhermit1, nonhermit2, nonhermit3, nonhermit4, nonhermit5, nonhermit6, nonhermit7, nonhermit8, nonhermit9, nonhermit10, nonhermit11, nonhermit12, nonhermit13}. Non-Hermitian physics is also pivotal in understanding the exotic properties of different systems \cite{sarisaman1, hamed2020, sarisaman2019, sarisaman20192}. This serves as the central motivation for our work. The integration of non-Hermitian frameworks with gravitational physics, such as gravitational waves, offers exciting opportunities for the development of this field with unprecedented results. 

Additionally, we will compare the value of the Hubble constant obtained using the non-Hermitian approach with well-known values from the literature and discuss the consistency of the solutions we have derived. Therefore, we immediately specify for the reader that, according to the analysis of data from the Planck satellite's Cosmic Microwave Background (CMB) observations, the value of the Hubble constant is $H_{0} = 67,4 \pm 0,5$ km s$^{-1}$ Mpc$^{-1}$\cite{Planckoll}. On the other hand, the SH0ES collaboration obtained a Hubble constant value of $H_{0} = 73,04 \pm 1,04$ km s$^{-1}$ Mpc$^{-1}$ by using Cepheid variables in a distance ladder to calibrate the data from Type Ia supernovae \cite{Shoes}. In addition to these, there is the Hubble constant obtained by combining a large number of datasets \cite{Valentino_2021_Clas}. It can be seen that these values fall roughly between the CMB and SHOES values\cite{Valentino_2021_Clas}. This is known Hubble tension problem. It is noteworthy that a substantial body of observational data and comprehensive analyses have been employed to accurately determine and refine the value of the Hubble parameter \cite{Di_Valentino_2021}. Finally, gravitational wave (GW) analyses have also been incorporated into these efforts. Several studies have attempted to constrain the Hubble parameter using GW data. For instance, values of $H_{0}\approx 70.0$ km s$^{-1}$ Mpc$^{-1}$ were obtained from GW170817 \cite{nature-2017a}, $H_{0}\approx 68.3$ km s$^{-1}$ Mpc$^{-1}$ from GW170817 combined with very long baseline interferometry (VLBI) measurements \cite{Mukherjee_2021}, $H_{0}\approx 73.4$ km s$^{-1}$ Mpc$^{-1}$ from the joint analysis of GW190521 and GW170817 \cite{arxiv-2020a}, and $H_{0}\approx 67.6$ km s$^{-1}$ Mpc$^{-1}$ from GW170817 combined with Zwicky Transient Facility (ZTF) data \cite{arxiv-2020b}.

In view of the cosmic history of the universe, CMB data correspond to the epoch of cosmic origins, while SHOES data pertain to a later period, specifically the era dominated by dark energy, known as the accelerated universe phase. Although the cause of this discrepancy, referred to as the Hubble tension \cite{Valentino_2021, Verde_2019}, has not yet been fully understood, this distinction is important for our study. This is because the evolutionary period of the universe from which the analyzed gravitational waves originated will be crucial. For instance, the neutron star merger GW170817, measured by the LIGO-Virgo collaboration, occurred 140 million light years away, in the accelerated universe era \cite{Abbott_2017}. We would like to emphasize that this point will be significant for our future considerations.

In our study, we identified spectral singularity points arising from non-Hermitian scattering of gravitational waves. We compared the frequencies of these waves with the known values of the Hubble constant and found a close match, highlighting the significance of non-Hermitian physics in the analysis of gravitational waves. Through this work, we aim to stimulate further interest in the applications of non-Hermitian physics to gravitational physics research and anticipate that it will lead to a deeper understanding of previously unexplored phenomena in this field.

This work is organized as follows. In Section \ref{S2}, we will introduce the perturbative Einstein field equations. The solutions to the perturbation equations will be obtained in terms of Bessel, Legendre, and Hankel polynomials in radial direction. Using the transfer matrix technique, we will derive the scattering amplitudes from spectral singularities. By exploring the relationship between these scattering amplitudes and the Hubble constant, we will discuss what gravitational waves represent within the framework of a non-Hermitian formalism. We will show that the results obtained within this non-Hermitian framework are consistent with the late-time accelerating universe. In Section \ref{S3}, we present some noteworthy conclusions.
    
\section{Scattering of Gravitational Waves: Non-Hermitian Perspective}
\label{S2}

\begin{figure}
    \begin{center}
    \includegraphics[scale=1]{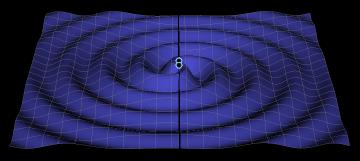}
    \caption{(Color online) Pictorial demonstration of Gravitational Wave production (from a binary system) and its propagation in space-time fabric. It highlights the structural analogy between gravitational waves and heliospheric current sheets, noting differences in properties and frequency ranges spanning from ultra-low to high frequencies as discussed in recent LIGO findings.}
    \label{fig1}
    \end{center}
    \end{figure}

\begin{figure}
    \includegraphics[scale=1]{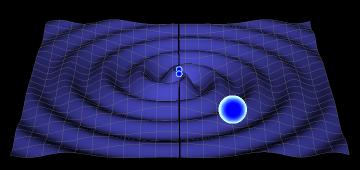}
    \caption{(Color online) Figure denotes the scattering configuration of the gravitational wave. Notice that the scatterer has almost a spherical shape.}
    \label{pl1}
    \end{figure}

In this study, we will consider a compact binary spiral collision representing the merger of two neutron stars. The propagation form of the gravitational waves emitted by such a collision can be shown in Fig.~(\ref{fig1}). This wave can be modeled as a wave-like movement of the space-time fabric and actually represents the propagation of the graviton particle at frequency $\omega$. This in fact occurs when the metric $g_{\mu\nu}$ is split into a background metric $\bar{g}_{\mu\nu}$ and a weak perturbative component $h_{\mu\nu}$ such that the gravitational wave propagates in that background metric $\bar{g}_{\mu\nu}$. Once the Einstein Equation is reduced to the first order in $h_{\mu\nu}$ by employing the TT (transverse-traceless) gauge, one obtains the following wave equation corresponding to the tensor perturbations $h_{\mu\nu}$
\be
\square{h}_{\mu\nu} + 
2\bar{R}_{\alpha\mu\nu\beta}h^{\alpha\beta} = \frac{\kappa^2}{4}(h_{\mu}^{\alpha}\bar{T}_{\alpha\nu} + h_{\nu}^{\alpha}\bar{T}_{\alpha\mu} + \frac{1}{2}\bar{g}_{\mu\nu}\bar{g}^{\alpha\beta}\delta^{(1)}T_{\alpha\beta} - 2\delta^{(1)}T_{\mu\nu})\label{303}.
\ee
Here the barred quantities $\bar{R}_{\alpha\mu\nu\beta}$, $\bar{T}_{\mu\nu}$ and $\bar{g}_{\mu\nu}$ correspond to the ones in the zeroth order equations in the $h_{\mu\nu}$ with background metric $\bar{g}_{\mu\nu}$ such that the associated Einstein's equation will be
\be
\bar{R}^{\mu\nu}-\frac{1}{2}\bar{g}^{\mu\nu}\bar{R} = \frac{\kappa^2}{4}\bar{T}^{\mu\nu}.
\ee
Also, $\delta^{(1)}T_{\mu\nu}$ represents the first order correction to the tensor field $T_{\mu\nu}$. For the cosmological considerations, we take $\bar{g}_{\mu\nu}$ to be the flat FLRW metric such that $\bar{g}_{00}=1$ and $\bar{g}_{ij}=-a^2\delta_{ij}$, where the gravitational wave propagates through the perfect fluid medium. Therefore, one obtains the gravitational wave
the equation in the FLRW universe as follows
\be
\Ddot{h}^{i}_{j} + 3\textit{H}\dot{h}^{i}_{j} - \frac{\partial^2}{a^2}h^{i}_{j} = 0,\label{grwaveeqn}
\ee
where $H := \dot{a} / a$ is the Hubble parameter, $h^{i}_{j} = \bar{g}^{ik}h_{kj}$ and $\partial^2 = \delta^{ij}\partial_{i}\partial_{j}$. Notice that the speed of gravitational wave $c_g$ is the coefficient of the $\frac{\partial^2}{a^2}$ term and in this case $c_g = 1$. See Appendix \textbf{A} and \textbf{B} for the details of the complete derivation of Eq.~(\ref{grwaveeqn}). In case of time-harmonic wave propagation, the form of the metric perturbation turns out to be $h^{i}_{j} (\vec{r}, t) = e^{-i \omega t} h^{i}_{j} (\vec{r})$ such that Eq.~(\ref{grwaveeqn}) is reduced to the following 3-dimensional Helmholtz equation
\be
\partial^2 h^{i}_{j} (\vec{r}) + \omega a^2 (\omega + 3i H) h^{i}_{j} (\vec{r}) = 0.\label{3dimhelm}
\ee
The solution of this equation in spherical coordinates gives rise to the following form in all regions of space
\be
h^{i}_{j} (r, \theta, \phi) = \left[ C \textrm{P}_{\ell}^{m} (\cos{\theta}) + D \textrm{Q}_{\ell}^{m} (\cos{\theta}) \right] e^{im\phi} \left\{\begin{array}{cc}
    \left[A_1 \textrm{h}_{\ell}^{(1)} (k_r r) + B_1 \textrm{h}_{\ell}^{(2)} (k_r r)\right] & {\rm for}~~~~ r \in (\infty, R],\\
    \left[A_2 \textrm{j}_{\ell} (k_r r) + B_2 \textrm{n}_{\ell} (k_r r)\right] & {\rm for}~~~~ r \in [0, R]
    \end{array}\right.
\ee
\UseRawInputEncoding where $A, B, C$ and $D$ are constants which will be determined from boundary conditions, $\textrm{j}_{\ell}$ and $\textrm{n}_{\ell}$ are the first and second kinds of spherical Bessel functions of order $\ell$ respectively,  $\textrm{h}_{\ell}^{(1)}$ and $\textrm{h}_{\ell}^{(2)}$ are the first and second kinds of spherical Hankel functions of order $\ell$ respectively, $\textrm{P}_{\ell}^{m}$ and $\textrm{Q}_{\ell}^{m}$ are the first and second kinds of associated Legendre polynomials respectively, and $m = 0, \pm 1, \pm 2, \ldots$ and $\ell = 0, 1, 2, \ldots$. Here the quantity $k_r$ is identified by $k_r := a \sqrt{\omega^2+3i\omega H}$. At this point, let's imagine that the scattering problem has a spherical configuration due to the cosmological character of gravitational waves, see Fig.~(\ref{pl1}).

If we consider that a gravitational wave propagating in a vacuum is scattered from a dense matter medium of radius $R$, shown in the Fig.~(\ref{pl1}), the relevant boundary conditions will occur as (i) $(h_{{\rm vacuum}})^{i}_{j} (\mathbf{z}_R) = (h_{{\rm matter}})^{i}_{j} (\mathbf{z}_R)$, and (ii) $(h_{{\rm vacuum}})^{i '}_{j} (\mathbf{z}_R) = (h_{{\rm matter}})^{i '}_{j} (\mathbf{z}_R)$, where $\mathbf{z} := k_r r$ and $\mathbf{z}_R := k_r R$, and a prime denotes derivative with respect to the parameter $\mathbf{z}$, i.e. $' := \frac{d}{d\mathbf{z}}$. Therefore, one obtains the following boundary conditions,
\begin{align}
A_1 \textrm{h}_{\ell}^{(1)} (\mathbf{z}_R) + B_1 \textrm{h}_{\ell}^{(2)} (\mathbf{z}_R) &= A_2 \textrm{j}_{\ell} (\mathbf{z}_R) + B_2 \textrm{n}_{\ell} (\mathbf{z}_R), \notag\\
A_1 \left[ \textrm{h}_{\ell}^{(1)}\right]' (\mathbf{z}_R) + B_1 \left[  \textrm{h}_{\ell}^{(2)}\right]' (\mathbf{z}_R) &= A_2 \textrm{j}_{\ell}' (\mathbf{z}_R) + B_2 \textrm{n}_{\ell}' (\mathbf{z}_R), \label{boundaryconds}
\end{align}
corresponding to each Bessel mode. These boundary conditions reveal the following transfer matrix construction, connecting the coefficients $A_j$ and $B_j$ for $j= 1, 2$
   \be
   \left(
           \begin{array}{c}
             A_2 \\
             B_2 \\
           \end{array}
   \right) = \mathbb{M} \left(
           \begin{array}{c}
             A_1 \\
             B_1 \\
           \end{array}
   \right), \notag
   \ee
We notice that the transfer matrix $\mathbb{M}$ can be decomposed into two distinctive pieces as $\mathbb{M} := \mathbb{V}^{-1} \mathbb{U}$, where the matrices $\mathbb{U}$ and $\mathbb{V}$ take the following forms
\be
\mathbb{U} = \left(
                       \begin{array}{cc}
                         \textrm{h}_{\ell}^{(1)} (\mathbf{z}_R) & \textrm{h}_{\ell}^{(2)} (\mathbf{z}_R) \\
                         \left[ \textrm{h}_{\ell}^{(1)}\right]' (\mathbf{z}_R) & \left[ \textrm{h}_{\ell}^{(2)}\right]' (\mathbf{z}_R) \\
                       \end{array}
                     \right),\qquad 
                     \mathbb{V} = \left(
                       \begin{array}{cc}
                         \textrm{j}_{\ell} (\mathbf{z}_R) & \textrm{n}_{\ell} (\mathbf{z}_R) \\
                         \textrm{j}_{\ell}' (\mathbf{z}_R) & \textrm{n}_{\ell}' (\mathbf{z}_R) \\
                       \end{array}
                     \right).
                     \label{transmatr}
\ee
Thus, one attains the transfer matrix manifestly in the following form,
\be
\mathbb{M} = \frac{1}{\textbf{W}[\textrm{j}_{\ell}, \textrm{n}_{\ell}] (\mathbf{z}_R)} \left(
                       \begin{array}{cc}
                        \textbf{W}[\textrm{h}_{\ell}^{(1)}, \textrm{n}_{\ell}] (\mathbf{z}_R) & \textbf{W}[\textrm{h}_{\ell}^{(2)}, \textrm{n}_{\ell}] (\mathbf{z}_R) \\
                          \textbf{W}[\textrm{j}_{\ell}, \textrm{h}_{\ell}^{(1)}] (\mathbf{z}_R) & \textbf{W}[\textrm{j}_{\ell}, \textrm{h}_{\ell}^{(2)}] (\mathbf{z}_R) \\
                       \end{array}
                     \right),
\ee
where $\textbf{W}[f, g] (\mathbf{z}) := f  \overset{\leftrightarrow}{\partial_{\mathbf{z}}} g = f (\mathbf{z}) \partial_{\mathbf{z}} g (\mathbf{z}) - g (\mathbf{z}) \partial_{\mathbf{z}} f (\mathbf{z})$ is the Wronskian of the differentiable functions $f (\mathbf{z})$ and $g (\mathbf{z})$. We require that zeros of $M_{22}$ components of the transfer matrix corresponding to real $k_r$ values give rise to spectral singularity points. This is provided if $\textbf{W}[\textrm{j}_{\ell}, \textrm{h}_{\ell}^{(2)}] (\mathbf{z}_R) = 0$, which yields
\be
\textrm{j}_{\ell} (\mathbf{z}_R) \left[ \textrm{h}_{\ell}^{(2)}\right]' (\mathbf{z}_R) =  \textrm{j}_{\ell}' (\mathbf{z}_R) \textrm{h}_{\ell}^{(2)} (\mathbf{z}_R). \label{eqss}
\ee
On the other hand, we recall the identity corresponding to spherical Bessel and Hankel functions of order $\ell$ as follows,
\be
\textrm{u}_{\ell}' (\mathbf{z}) = \frac{d \textrm{u}_{\ell} (\mathbf{z})}{d \mathbf{z}} := \frac{\ell \textrm{u}_{\ell -1} (\mathbf{z}) -(\ell + 1)\textrm{u}_{\ell + 1} (\mathbf{z})}{2\ell + 1}, \label{eqderid}
\ee
where $\textrm{u}_{\ell}$ represents the spherical Bessel or Hankel type functions of order $\ell$. Since the explicit form of these equations is not easily obtained, we inquire about their asymptotic expressions that are valid for $\mathbf{z} >> 1$. For typical situations, the radius of the scattering spherical object is much larger than the wavelength of the wave. Therefore, for practical purposes that we will consider, the condition $\mathbf{z} >> 1$ holds and the asymptotic treatment of (\ref{eqss}) and (\ref{eqderid}) provides extremely reliable results.

The asymptotic expansion of the spherical Bessel and Hankel functions, $\textrm{j}_{\ell}$ and $\textrm{h}_{\ell}^{(2)}$, that are valid in the large argument limit ($\mathbf{z} >> 1$), have the form
\begin{align}
\textrm{j}_{\ell} (\mathbf{z}) &= \frac{\sin{(\mathbf{z} -\frac{\pi \ell}{2}})}{\mathbf{z}} \sum_{s=0}^{\infty} \frac{(-1)^s \mathcal{A}_{2s}(\ell)}{\mathbf{z}^{2s}} + \frac{\cos{(\mathbf{z} -\frac{\pi \ell}{2}})}{\mathbf{z}} \sum_{s=0}^{\infty} \frac{(-1)^s \mathcal{A}_{2s + 1}(\ell)}{\mathbf{z}^{2s + 1}}, \notag\\
\textrm{h}_{\ell}^{(2)} (\mathbf{z}) &= \frac{e^{-i(\mathbf{z} -\frac{\pi \ell}{2}})}{\mathbf{z}} \left[ i \sum_{s=0}^{\infty} \frac{(-1)^s \mathcal{A}_{2s}(\ell)}{\mathbf{z}^{2s}} +  \sum_{s=0}^{\infty} \frac{(-1)^s \mathcal{A}_{2s + 1}(\ell)}{\mathbf{z}^{2s + 1}} \right ], \label{expansions}
\end{align}
where 
\be
\mathcal{A}_{k}(\ell) := \frac{\Gamma (\ell + k + 1)}{2^k k! \Gamma (\ell - k + 1)} = \frac{\prod_{j=0}^{2k -1} (\ell + k - j)}{2^k k!},\label{Aks}
\ee
and $\Gamma$ stands for the Gamma function. In the asymptotic limit $\mathbf{z} >> 1$, instead of considering all the terms in the series expansions given in the equations (\ref{expansions}), it is essential to have enough terms for our purposes. Thus, we will consider the terms with the coefficients $\mathcal{A}_{0}, \mathcal{A}_{1}$ and $\mathcal{A}_{2}$. Contributions of other terms can be neglected in light of the specified large argument limit. Using the expression given in (\ref{Aks}), we can specify these coefficients as
\be
\mathcal{A}_{0}(\ell) = 1,\qquad \mathcal{A}_{1}(\ell) = \frac{\ell (\ell + 1)}{2}, \qquad \mathcal{A}_{2}(\ell) = \frac{(\ell + 2)(\ell + 1) \ell (\ell - 1)}{8}. \notag
\ee
Substituting these coefficients in the expression (\ref{expansions}), then using the equations (\ref{eqss}) and (\ref{eqderid}), we obtain the result up to the order of $\mathcal{O} (\mathbf{z}_R^{-3})$ given below, 
\be
\tan (\mathbf{z}_R -\frac{\pi \ell}{2}) = \zeta_{\ell} (\mathbf{z}_R). \label{eq313}
\ee
The function $\zeta_{\ell} (\mathbf{z}_R )$ specified here is a complex-valued expression defined by a continuous map $\zeta_{\ell}: \mathbb{C} \rightarrow \mathbb{C}$, and is defined explicitly as below,
\be
\zeta_{\ell} (\mathbf{z}_R) := \frac{i \textit{g}_{\ell}(\mathbf{z}_R) \textit{h}_{\ell}(\mathbf{z}_R) - \mathcal{A}_{1}(\ell) \textit{f}_{\ell}(\mathbf{z}_R)}{\textit{f}_{\ell}(\mathbf{z}_R) [\mathbf{z}_R^2 - \mathcal{A}_{2}(\ell)]-i\mathbf{z}_R (2\ell + 1)\textit{h}_{\ell}(\mathbf{z}_R)}.
\ee
Here, identifications $\textit{f}_{\ell}$, $\textit{g}_{\ell}$ and $\textit{h}_{\ell}$ employed are defined as
\begin{align}
\textit{f}_{\ell}(\mathbf{z}_R) &:= 8 (2\ell + 1) \mathbf{z}_R^2 + 8 i (\ell + 1)(2\ell + 1)\mathbf{z}_R +(\ell^5-3\ell^4-8\ell^3-15\ell^2+17\ell+6),\notag\\
\textit{g}_{\ell}(\mathbf{z}_R) &:=(2\ell + 1) [\mathbf{z}_R^2 + 4\ell(\ell + 1)],\notag\\
\textit{h}_{\ell}(\mathbf{z}_R) &:= 8 \mathbf{z}_R^2 - 4i\mathbf{z}_R (\ell + 1)(\ell + 2)-(\ell + 2)(\ell + 1) \ell (\ell - 1).\notag
\end{align}
In view of these definitions, the expression (\ref{eq313}) yields the following result,
\be
\tan^{-1} [\zeta_{\ell} (\mathbf{z}_R)] = \mathbf{z}_R -\frac{\pi \ell}{2}. \label{eq315}
\ee
At this point, we can use the well-known identity for $\tan^{-1}$ given in the form
\be
\tan^{-1} (\mathit{z}) := \pi m + \frac{1}{2 i} \ln{\left(\frac{1 + i\mathit{z}}{1 - i \mathit{z}}\right)}. \notag
\ee
Here $m$ takes positive or negative integer values, $m = 0, \pm 1, \pm 2, \ldots$. Thus, we can describe (\ref{eq315}) as follows: 
\be
\mathbf{z}_R -\frac{\pi \ell}{2} = \pi m + \frac{1}{2} \ln{\left(\frac{1 + i\zeta_{\ell} (\mathbf{z}_R)}{1 - i \zeta_{\ell} (\mathbf{z}_R)}\right)}.
\ee
This is a complex expression and can be analyzed by decomposing it into its real and imaginary parts. For this purpose, we can write the real and imaginary parts of the quantities $\mathbf{z}_R$ and $\zeta_{\ell}$ as follows,
\begin{align}
\mathbf{z}_R := \mathbf{z}_{R, r} + i \mathbf{z}_{R, i}, \qquad
\zeta_{\ell} := \zeta_{\ell, r} + i \zeta_{\ell, i}. \notag
\end{align}
Since it is known that $\mathbf{z}_R = a R \sqrt{\omega^2 + 3 i\omega H}$, it is possible to obtain the real and imaginary parts correspondingly. Since we focus here on solutions corresponding to real values of the scale parameter $a$, we obtain the following results in the real scale factor,
\begin{align}
\mathbf{z}_{R, r} &= \pi (m + \frac{\ell}{2}) + \frac{1}{2}\ln{\left(\frac{\sqrt{4\zeta_{\ell, r}^2 + \left(1 -\left|\zeta_{\ell}\right|^2\right)^2}}{\zeta_{\ell, r}^2 + (1 + \zeta_{\ell, i})^2}\right)}, \label{zr}\\
\mathbf{z}_{R, i} &= \frac{1}{2} \tan^{-1} \left(\frac{2\zeta_{\ell, r}}{1-\left|\zeta_{\ell}\right|^2}\right). \label{zi}
\end{align}

These two equations describe the non-Hermitian scattering behavior of gravitational waves, leading to the formation of spectral singularities. The system's non-Hermitian scattering properties can be controlled by adjusting the parameters that appear in these equations. This can be done by taking advantage of the known characteristics of gravitational waves and the specific parameters of the scattering system. A crucial parameter to determine in this context is the scale factor $a(t)$, which continues to be a subject of active research and precise measurement. While the scale factor is typically assessed using real values in the literature, complex values are also considered in some cases. To validate the consistency of our method, we will begin by using the known value of the scale factor in a universe dominated by dark energy. To this end, the scale factor in a FLRW universe primarily dominated by dark energy is given by \cite{aydiner}.

The scale factor relates the distances between objects in the universe at different times. The scale factor increases over time in an expanding universe. For instance, if $a(t)$ was smaller in the past, it indicates that the universe was more compact. The expansion is often described by the Hubble parameter, which is related to the rate of change of the scale factor. Different cosmological models predict different behaviors for the scale factor. For example, in a matter-dominated universe, the scale factor increases as $t^{2/3}$, which arises from the Friedmann solution to Einstein's field equations. On the other hand, in a universe dominated by dark energy (like a cosmological constant), the scale factor behaves as $e^{H t}$, which is a result of the de Sitter solution for the case where $k=0$. These solutions are consistent with observations\footnote{The significant discrepancy in the universe's mass-energy content suggests the presence of an unknown component, often modeled as a cosmological constant or dark energy. This deficit, first noted by Einstein, underscores both the limitations of current gravitational models and the unresolved nature of gravity itself.}.

However, the observational result for the scale factor remains debated. Since the scale factor is determined by the Hubble parameter, knowing its exact value is crucial. As emphasized in the introduction, there are two values of the Hubble parameter that are widely accepted in the literature: One of these values is 
$H_{0} = 67,4 \pm 0,5$ km s$^{-1}$ Mpc$^{-1}$, obtained from CMB data by the Planck satellite, while the other is $H_{0} = 73,04 \pm 1,04$ km s$^{-1}$ Mpc$^{-1}$, derived from data on Type Ia supernovae. \footnote{GW data and their corresponding Hubble constant $H_0$ values are also available, and the discussion regarding these has already been provided in the introduction section. Please refer to that section for details.}
These values are observational and consist of real numbers. However, the main issue we would like to highlight here is how the scale factor is derived from a theoretical model. As far as is known in the literature, a scale factor that accurately represents the entire history of the universe in a functional or mathematical form has not yet been achieved. In a recent study we are involved with, however, a hybrid scale factor that accurately represents both the dominant matter era and the dark energy-dominated era has been analytically derived, based on the interaction between dark matter and dark energy.
\begin{eqnarray} \label{Model-NewS}
	a(t) = a_{0} \left(\frac{t}{t_{0}}\right)^{\alpha}  e^{h\frac{t}{t_{0}}} \ ,
\end{eqnarray}
where certain parameters in this expression are $\alpha=2/3$, $h\sim 0.73$. In this hybrid scaling factor model, the parameter $\alpha=2/3$ corresponds to the matter-dominated era, which is consistent with the FLRW solution. On the other hand, the small $h$ represents the Hubble constant $H_{0} = 73,04 \pm 1,04$ km s$^{-1}$ Mpc$^{-1}$ as an upper bound \cite{aydiner} \footnote{Notice also the lower bound $H_{0} = 67,4 \pm 0.5$ km s$^{-1}$ Mpc$^{-1}$}. Since we are particularly interested in the dark energy-dominated era, we will focus on the Hubble parameter, which appears in the exponential term $e^{H t}$. If the Hubble parameter \cite{hubble1, hubble2, hubble3, hubble4} is taken as $H := \dot{a}(t) / a (t)$, then $H \approx H_0$. The currently known and accepted value of the Hubble constant in a FLRW universe in the literature is as follows \cite{aydiner, hubble5, hubble6}
\be
H_0 = \sqrt{8\pi G \rho_{full}/3} = \sqrt{\Lambda /3}.
\ee
Here $\Lambda$ is the cosmological constant as known, and if it is fixed to a positive value, it can be seen that the geometry of our universe will be equivalent to de Sitter universe \cite{desitter1, desitter2, desitter3, desitter4}. In this case, the most specifically measured value of the cosmological constant is $\Lambda \approx 2\times 10^{-35}~\textrm{s}^{-2}$. This gives the value of the Hubble constant as approximately $H_0 \approx 70.88~\textrm{km}~\textrm{s}^{-1} \textrm{Mpc}^{-1}$. This corresponds to approximately 13.79 billion years in Hubble time \cite{hubbleconst, hubbleconst2, hubbleconst3, hubbleconst4}.

However, the main focus of our study here is not on the scale factors found or proposed in the literature. Instead, we are in search of more unconventional scale factors that can allow us to test our model. In this work, we discuss how gravitational waves would scatter if the scale factor were a complex value. We require this framework to analyze the scattering of gravitational waves within a non-Hermitian framework. Fortunately, there are strong motivations for this choice in the existing literature. It is well-known that complex or complex analytic functions are frequently used in gravitational theories. For instance, in many applications, the analytic continuation of a real analytic manifold is transformed into a complex form to obtain a complex spacetime \cite{Newman_1988}. As examples, open and closed Friedmann models, de Sitter and anti-de Sitter spacetimes, as well as Kerr and Schwarzschild metrics can all be related to complex forms  \cite{Hartle_Hawking}. Another significant example of a complex spacetime approach in quantum cosmology is Hartle-Hawking’s "no boundary condition" proposal. Similarly, there are complex metric models that change the signature of the metric \cite{Ellis_1992, Greensite_1993, Witten_comp}. Considering complex spacetimes, the Wheeler-DeWitt model in quantum cosmology is another relevant example  \cite{DeWitt_1967}. However, the most fundamental approach within the FRLW framework involves choosing the scale factor to be complex. This choice is made quite simply as follows\cite{John_1996,John_1997,John_1999}:
\begin{eqnarray} \label{scale-in}
	\hat{a}(t) = a(t) e^{i\beta(t)} = a_{0} e^{\alpha(t) + i \beta(t)}
\end{eqnarray}
Here, $a_{0}$ is a constant and represents the value of the scale factor at $t=0$. This choice allows us to express the scale factor and the Hubble parameter in complex form. The square of the expression in Eq. (\ref{scale-in}) is
\begin{eqnarray} \label{scale-in-square}
	a^{2} (t) = \mid \hat{a}(t)\mid^{2}  =  a^{2}_{0} + t^{2}
\end{eqnarray}
such that it takes the form of the scale factor present in the FRLW metric. In quantum cosmology calculations, the scale constant is given by $a_{0} = \sqrt{2G/3 \pi } \approx l_{pl}$, which is the Planck length. This result shows that the real component of the complex scale factor remains real in the observable space. However, the complex component plays an important role in scattering processes.

\section{Results and Discussion}
\label{S3}

Measuring gravitational waves is a very difficult task and this can only be done using very sensitive detectors. Since the gravitational waves reaching our Earth have very small amplitudes, in order to detect them, all other signals in the environment, including the smallest sound and noise signals, must be eliminated. As a result of all this, the measurable frequency values of gravitational waves depend on the source from which the gravitational waves originate. Gravitational waves generated by a binary system of uncharged black holes lie at the lower end of the gravitational wave spectrum, ranging from $10^{-7}$ to $10^5$ Hz. An astrophysical source at the high frequency end of the gravitational wave spectrum (above $10^5$ Hz, potentially reaching up to $10^{10}$ Hz) is believed to produce relic gravitational waves, which are thought to be faint traces of the Big Bang, similar to the cosmic microwave background \cite{highfreq}. At these high frequencies, there is a potential for sources to be "artificial," referring to gravitational waves that are generated and observed in a laboratory environment \cite{highfreq2, highfreq3}. The Nobel Lecture "LIGO and Gravitational Waves III" comprehensively addresses the anticipated future scope of gravitational wave detection, spanning multiple frequency domains, including high-frequency, low-frequency, very-low-frequency, and ultra-low-frequency bands. Furthermore, it emphasizes the open question regarding whether gravitational waves may encompass a frequency spectrum broader than currently recognized \cite{thorne2018nobel}.

Typically, we consider that the measurable gravitational wave frequencies are in the range of $10^{-16}~\textrm{Hz} < f < 10^{4}~\textrm{Hz}$ \footnote{Please note the relation between $f$ and $\omega$, $\omega = 2\pi f$.} \cite{freq1, freq2, freq3, freq4, freq5, freq6, freq7, freq8, freq9}. While the lower limit is generally formed by gravitational waves that formed in the early times of the universe and can still be detected, the values close to the upper limit are waves that occurred closer to our world and in later times. Our goal is to confirm that the method we employed and the results we obtained are consistent with these data and observations. In addition, this method allows for obtaining information about the sources from which gravitational waves originate and investigating their other properties. Since the approach we use is quite new and points to a new aspect of non-Hermitian physics, it can be developed and contribute to the understanding of many other unknown secrets investigated in gravitational wave physics.

\begin{figure}
\centering
\begin{tikzpicture} 
        \node[anchor=north west,inner sep=0pt] at (0,0){\includegraphics[scale=0.6]{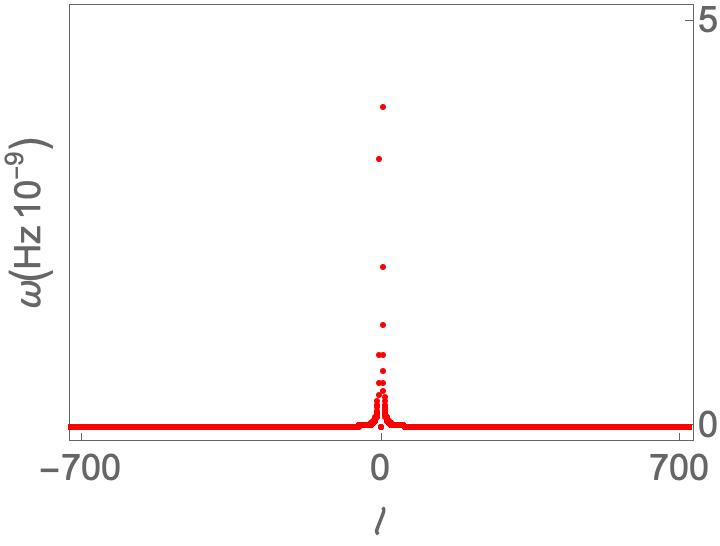}};
        \node[font=\sffamily\bfseries\large] at (9ex,-2.5ex) {(a)}; 
        \end{tikzpicture} 
        \begin{tikzpicture} 
        \node[anchor=north west,inner sep=0pt] at (0,0){\includegraphics[scale=0.6]{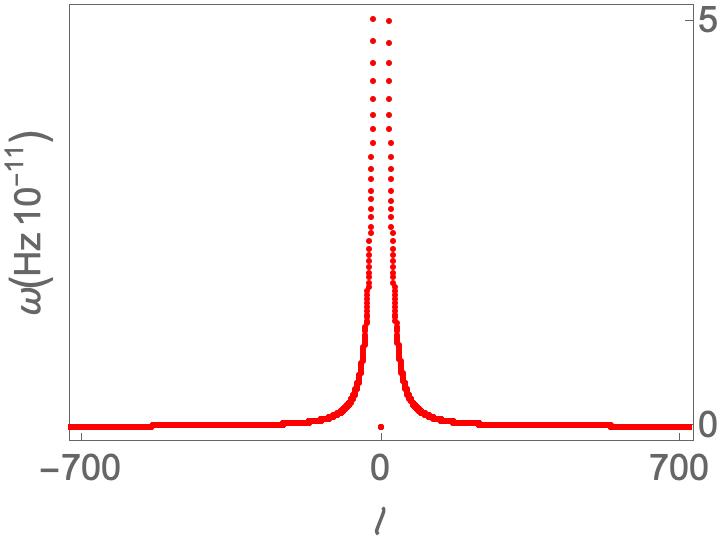}};
        \node[font=\sffamily\bfseries\large] at (9ex,-2.5ex) {(b)}; 
        \end{tikzpicture}\\
                \begin{tikzpicture} 
        \node[anchor=north west,inner sep=0pt] at (0,0){\includegraphics[scale=0.6]{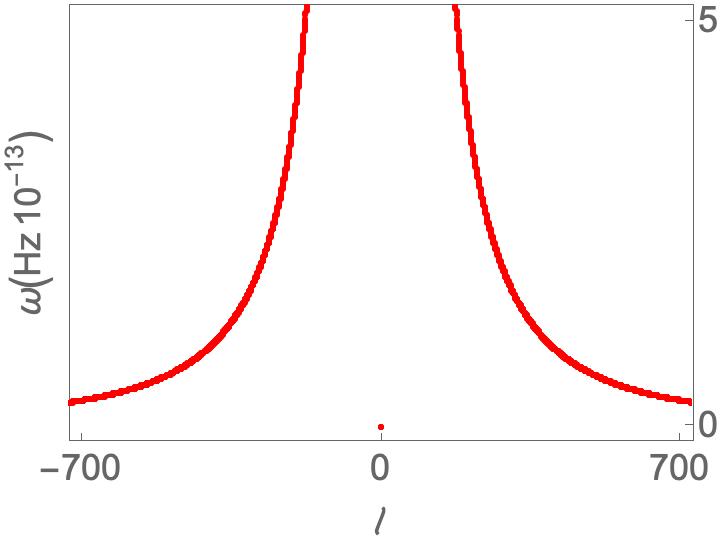}};
        \node[font=\sffamily\bfseries\large] at (9ex,-2.5ex) {(c)}; 
        \end{tikzpicture}
        \begin{tikzpicture} 
        \node[anchor=north west,inner sep=0pt] at (0,0){\includegraphics[scale=0.6]{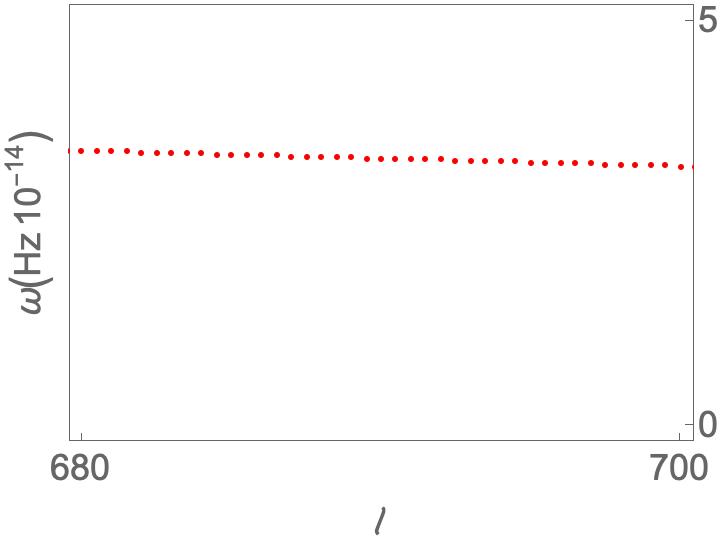}};
        \node[font=\sffamily\bfseries\large] at (9ex,-2.5ex) {(d)}; 
        \end{tikzpicture}
    \caption{(Color online) The figures illustrate the locations of the spectral singularity points, derived from the real zeros of the F and G functions in the $\ell$-$\omega$-plane. Panel (a) provides an overall view of the spectral singularity points, while the subsequent panels offer a more detailed, scaled view of the remaining points. It is evident that for certain Bessel orders near zero, no spectral singularity points exist, and in this region, the gravitational wave frequencies attain their maximum values. The Hubble constant $H_0 = 70.88~\textrm{km}~\textrm{s}^{-1} \textrm{Mpc}^{-1}$, is used throughout this analysis.}
    \label{wl}
\end{figure}
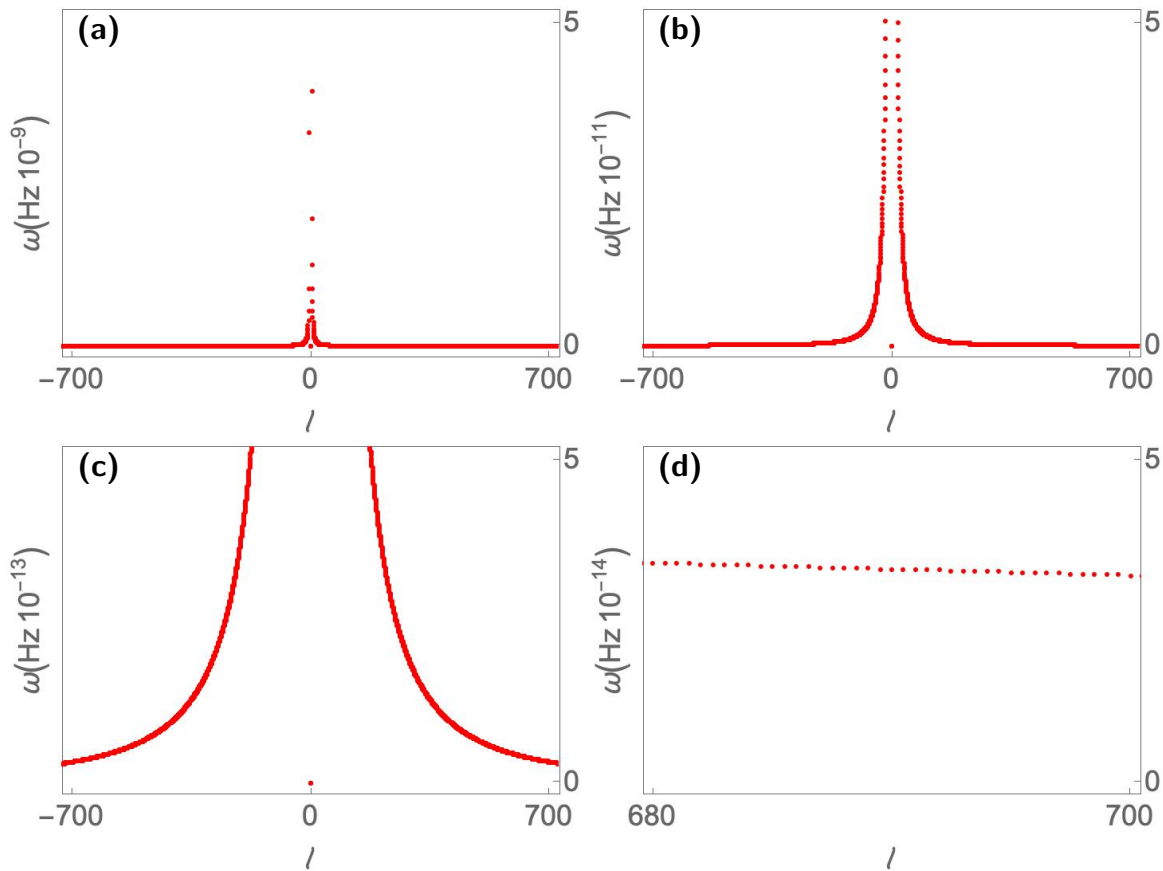

To determine the optimal parameters leading to spectral singularities, we begin by using the spectral singularity conditions from Eqns.~(\ref{zr}) and (\ref{zi}). We introduce the following definitions,
\begin{align}
\mathbf{F} (m, \ell, \omega, H_0) &:= \mathbf{z}_{R, r} - \pi (m + \frac{\ell}{2}) - \frac{1}{2}\ln{\left(\frac{\sqrt{4\zeta_{\ell, r}^2 + \left(1 -\left|\zeta_{\ell}\right|^2\right)^2}}{\zeta_{\ell, r}^2 + (1 + \zeta_{\ell, i})^2}\right)}, \label{zr2}\\
\mathbf{G} (m, \ell, \omega, H_0) &:= \mathbf{z}_{R, i} - \frac{1}{2} \tan^{-1} \left(\frac{2\zeta_{\ell, r}}{1-\left|\zeta_{\ell}\right|^2}\right), \label{zi2}
\end{align}
together with the identifications,
\begin{align} 
m&=25128,~~ \ell=12564,~~ R\approx 2.5 \times 10^{4}~\textrm{km}, 
\end{align}
such that these parameters correspond to the gravitational waves generated in the early times of the universe (after the cosmic beginning), which have been measured here on Earth. Given that there are several parameters to determine, we begin by calculating the Bessel orders and their corresponding gravitational wave frequencies for various mode numbers, using the known value of the Hubble constant $H_0 \approx 70.88~\textrm{km}~\textrm{s}^{-1} \textrm{Mpc}^{-1}$. To do this, we examine the real zeros of the $\mathbf{F}$ and $\mathbf{G}$ functions. The resulting values are presented in Fig.~\ref{wl}.
As seen in these graphs, the gravitational wave frequencies reach their highest values at the lowest Bessel mode order values. Interestingly, spectral singularity points do not appear at $\ell = -3, -2, -1, 0, 2$. Additionally, gravitational wave frequencies can be observed at other integer Bessel mode orders. The resulting pattern forms a pipe-like structure, revealing a distribution centered around $\ell = 0$. To better interpret these graphs, note that the vertical scale in each panel is reduced by a factor of 100. In panel d, you can see that the spectral singularity points decrease consistently with Bessel order $\ell$.

Next, it is crucial to verify whether the functions $\mathbf{F}$ and $\mathbf{G}$ produce real zeros at the observed gravitational wave frequencies and the corresponding values of the Hubble constant. The graphs of the functions $\mathbf{F}$ and $\mathbf{G}$ with respect to $\omega$ and $H_0$, shown below, show a strong agreement with the measured values of the parameters $\omega$ and $H_0$ at the zeros of these functions.

\begin{figure}
\centering
\includegraphics[scale=0.6]{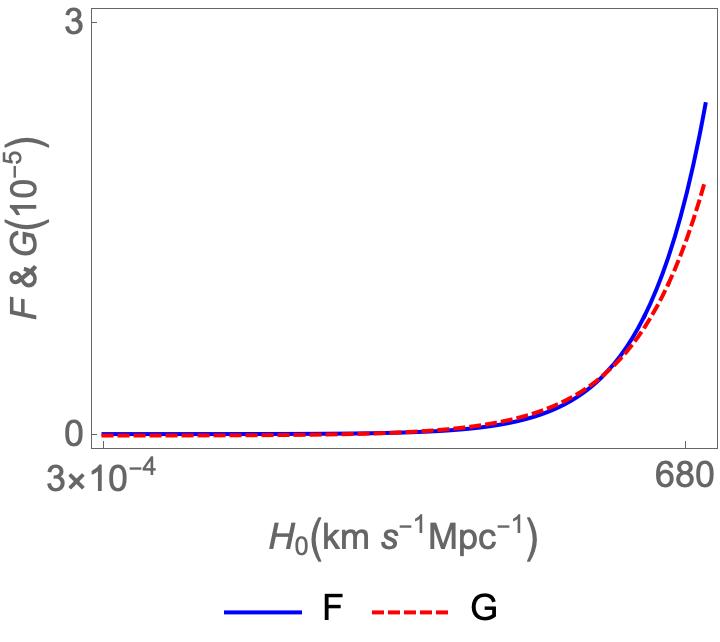}
\includegraphics[scale=0.6]{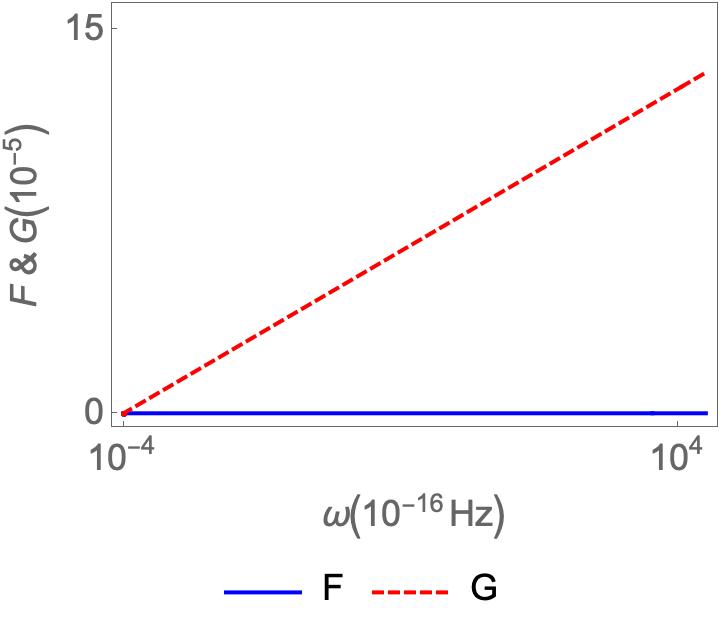}
    \caption{(Color online) The figures display the graphs of the functions $\mathbf{F}$ and $\mathbf{G}$ as functions of the Hubble constant $H_0$ and the gravitational wave frequency $\omega$. The known values of these parameters correspond to the zeros of the $\mathbf{F}$ and $\mathbf{G}$ functions within the measured value ranges. In the left panel, the gravitational wave frequency is set to $\omega = 10^{-16}~Hz$, while in the right panel, the Hubble constant is set to $H_0 = 70.88~\textrm{km}~\textrm{s}^{-1} \textrm{Mpc}^{-1}$.}
    \label{wh}
\end{figure}

These graphs provide insights into the possible ranges for the $m$ and $\ell$ values of the spectral singularity points. Accordingly, the spectral singularities at different $m$ values corresponding to the Bessel modes $\ell = 12564, 20$ and $3740$ are listed in the Table~\ref{table1}. This table indicates the frequencies of gravitational waves to be detected and Hubble constant values at which the spectral singularities occur.

\begin{table*}[ht]
\centering 
\begin{tabular}{| c | c |c |c |} 
\hline 
~~$\textcolor{blue}{\bf \#~of~SS}$ ~~ &~~$\textcolor{blue}{\ell}$~~ & ~~$\textcolor{blue}{\bf \omega~[Frequency~of~Gravitational~Waves]}$ ~~& ~~$\textcolor{blue}{\bf H_0~[Hubble~Constant]}$ ~~\\ [0.5ex] 
\hline \hline & & &\\
$1$ &~~~~12564~~~& ~~~~$10^{-22}~\textrm{Hz}$~~~ &~~0.704~\textrm{Km}~\textrm{s}$^{-1}$\textrm{Mpc}$^{-1}$\\
& & &\\
\hline & & &\\ 
$2$ &~~~~12564~~~~& ~~~~$10^{-21}~\textrm{Hz}$~~~&~~0.701~\textrm{Km}~\textrm{s}$^{-1}$\textrm{Mpc}$^{-1}$\\
& & &\\
\hline & & &\\ 
~~$3$~~ &~~~~12564~~~~& ~~~~$10^{-16}~\textrm{Hz}$~~~&~~69.98~\textrm{Km}~\textrm{s}$^{-1}$\textrm{Mpc}$^{-1}$\\
& & &\\
\hline & & &\\ 
$4$ &~~~~12564~~~~& ~~~~$10^{-16}~\textrm{Hz}$~~~ &~~0.698~\textrm{Km}~\textrm{s}$^{-1}$\textrm{Mpc}$^{-1}$\\
& & &\\
\hline & & &\\ 
$1$ &~~~~20~~~~& ~~~~$10^{-23}~\textrm{Hz}$~~~ &~~0.710~\textrm{Km}~\textrm{s}$^{-1}$\textrm{Mpc}$^{-1}$\\
& & &\\
\hline & & &\\ 
$2$ &~~~~20~~~~& ~~~~$10^{-14}~\textrm{Hz}$~~~ &~~70.03~\textrm{Km}~\textrm{s}$^{-1}$\textrm{Mpc}$^{-1}$\\
& & &\\
\hline & & &\\ 
$1$ &~~~~3740~~~~& ~~~~$10^{-23}~\textrm{Hz}$~~~ &~~0.694~\textrm{Km}~\textrm{s}$^{-1}$\textrm{Mpc}$^{-1}$\\
& & &\\
\hline & & &\\ 
$2$ &~~~~3740~~~~& ~~~~$10^{-12}~\textrm{Hz}$~~~ &~~70.41~\textrm{Km}~\textrm{s}$^{-1}$\textrm{Mpc}$^{-1}$\\
& & &\\
\hline 
\end{tabular}
\caption{Table shows Frequency and Hubble constant values corresponding to the real zeros of the $\mathbf{F}$ and $\mathbf{G}$ functions for different Bessel mode orders. These spectral singularity points are associated with stable gravitational wave configurations.} 
\label{table1} 
\end{table*}

\begin{figure*}
\centering
\includegraphics[scale=0.8]{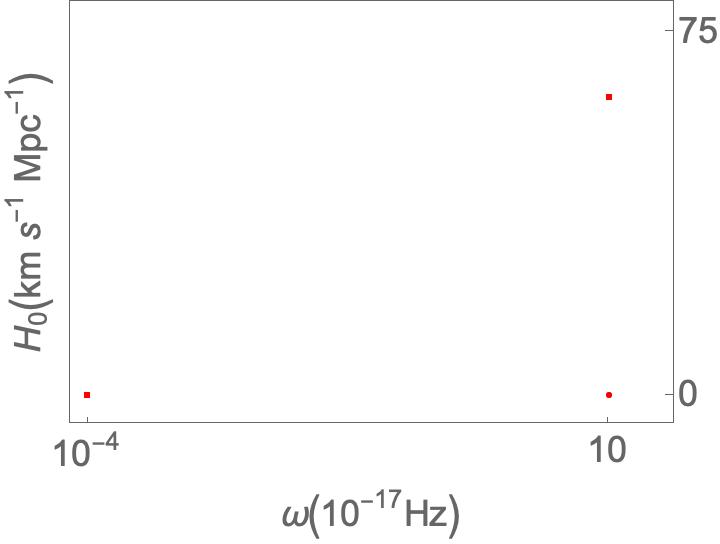}
    \caption{(Color online) The figure shows the spectral singularity points in the $\omega$-$H_0$ plane. Here the Bessel mode order is taken as $\ell = 12564$. There exist 4 spectral singularity points in the figure, one of which is very close to the point at the bottom left and cannot be noticed on the scale shown. The existence of this point can be seen in Table~\ref{table1}. }
    \label{wh2}
\end{figure*}

As shown in Table~\ref{table1} and Fig.~\ref{wh2}, the Bessel function orders corresponding to detectable gravitational wave frequencies from Earth are limited, with only a few discrete spectral points in these orders. These points provide the uniquely determined values for the Hubble constant and the gravitational wave frequencies scattered from Earth. If gravitational waves scattered by a cosmological object, other than the Earth, can be detected, it would be possible to determine their frequency and Bessel mode order using a similar approach. However, given the current uncertainty about how such waves could be measured, we will not delve into speculative calculations on this matter.

The results we obtained highlight the impact of non-Hermitian physics methods on gravitational wave scattering, demonstrating the effectiveness of this approach. Given that this method has proven successful in numerous electromagnetic problems, its strong performance with gravitational waves could offer valuable insights into various aspects of their nature. This serves as a clear indication of the non-Hermitian character of gravitational waves.

\section{Concluding Remarks}
\label{S4}

In conclusion, this study introduces a novel non-Hermitian approach to the scattering of gravitational waves, marking the first exploration of this phenomenon within the framework of non-Hermitian physics. By deriving complex potential solutions to the Einstein field equations, we examined the scattering of gravitational waves by round-like cosmic objects. Our approach enabled the identification of exceptional points unique to non-Hermitian systems, leveraging the intriguing characteristics of the transfer matrix. This work lays the foundation for further investigations into the non-Hermitian properties of detected gravitational waves, offering new insights into their behavior and interactions.

Inspired by scattering theory, we explore the fundamentals of identifying spectral singularities using the transfer matrix approach, such that boundary conditions are highly effective~\cite{prl-2009, CPA, pra-2017a, cpa3}. By using the well-known characteristics of gravitational waves detected on Earth, we utilize the appropriate Bessel function orders and the Hubble constant value $H_0$ to verify our results against established findings in the literature. We derived the condition for spectral singularity in the equation corresponding to the real $k_r$ values, as given by Eqn.~\ref{eqss} The reason $k_r$ is real in this context is that unstable states can arise in the complex potential scenario within non-Hermitian physics. Since stable states are only found for real $k_r$ values, the spectral singularities associated with these stable states have been identified. These are presented in expressions (\ref{zr}) and (\ref{zi}).

The spectral singularity conditions we derived provide the optimal relationships between various parameters, enabling us to determine their ranges and probable values based on the known properties of gravitational waves. As demonstrated in Fig.~\ref{wl}, we successfully identified the detectable frequencies of gravitational waves and their corresponding Bessel mode orders, using the known value of the Hubble constant $H_0$. The results are in excellent agreement with the gravitational wave frequencies measured on Earth. Similarly, in Fig.~\ref{wh}, we showed that the spectral singularity conditions are fully satisfied for the given Bessel mode orders, measured frequencies, and the Hubble constant. Furthermore, our analysis allowed us to predict the gravitational wave frequencies for different Bessel mode orders and estimate the probable value of the Hubble constant, with the results summarized in Table~\ref{table1}. These findings underscore the strength of the method we employed and provide compelling evidence that gravitational waves exhibit non-Hermitian characteristics.

It is evident that distinct Bessel mode orders generate a range of frequencies that are closely spaced. While the cause of this phenomenon may not have been clear when these frequencies were first measured, our study reveals that it is a result of the non-Hermitian properties involved. Our study demonstrates that exploring and understanding the impact of non-Hermitian physics in gravitational wave research and in gravitational physics more broadly, will significantly advance our understanding of many unresolved issues in this field. In this respect, our study offers a new perspective on gravitational physics and presents a novel approach for researchers in this field.

Finally, the formalism presented in this study offers a novel framework through which various cosmological models can be revisited. Notably, it enables the derivation of alternative values for the Hubble constant that have not been previously predicted. Approaches such as Conformal Cyclic Cosmology and the Big Bang Theory, in particular, may benefit from re-examination under this new perspective, guided by the insights gained from our findings.\\[6pt]


\section*{Appendix}\label{grdetails}

\subsection{Graviton Propagation in a Medium}\label{medium}

In this appendix, we clearly demonstrate the graviton propagation equations in an arbitrary medium. We start with Einstein's equation
\begin{equation}
{R}^{\mu\nu} - \frac{1}{2}{g}^{\mu\nu}R = \frac{\kappa^2}{4}{T}^{\mu\nu}\label{einseq}
\end{equation}
we expand the metric ${g}^{\mu\nu} = \bar {g}_{\mu\nu} + {h}_{\mu\nu}$ where $\bar {g}_{\mu\nu}$ is the background metric and ${h}_{\mu\nu}$ is the weak perturbative component such that the gravitational wave propagates in that background. The stress tensor of a general matter field is defined as
\begin{equation}
    {T}^{\mu\nu} = \frac{-2}{\sqrt{-g}}\frac{\delta(\sqrt{-g}\cL_m)}{\delta{g}_{\mu\nu}}
\end{equation}
The equations of motion corresponding to the weak perturbative tensor field ${h}_{\mu\nu}$ can be obtained by expanding first ${G}^{\mu\nu} :={R}^{\mu\nu}-\frac{1}{2}{g}^{\mu\nu}R = \frac{\kappa^2}{4}{T}^{\mu\nu}$ to the linear order in ${h}_{\mu\nu}$ around the background metric $\bar {g}_{\mu\nu}$. The zeroth order equation in ${h}_{\mu\nu}$ leads to the background equation as follows
\begin{equation}
    \bar {R}^{\mu\nu} - \frac{1}{2}\bar {g}^{\mu\nu}\bar R = \frac{\kappa^2}{4}\bar {T}^{\mu\nu} 
\end{equation}
The first order corrections in ${h}_{\mu\nu}$ of (\ref{einseq}) can be obtained as follows
\begin{equation}
    \delta^{(1)}{R}^{\mu\nu} - \frac{1}{2}\delta^{(1)}{g}^{\mu\nu}\bar R - \frac{1}{2}\bar {g}^{\mu\nu}\delta^{(1)}R = \frac{\kappa^2}{4}\delta^{(1)}{T}^{\mu\nu}\label{20}
\end{equation}
Notice also, the first order connection to the Christoffel connection is yielded by
\begin{equation}
\begin{split}
    \delta^{(1)}\Gamma^{\mu}_{\nu\alpha} &= - \frac{1}{2}{h}^{\mu\beta}(\partial_{\nu}\bar {g}_{\alpha\beta} + \partial_{\alpha}\bar {g}_{\nu\beta} - \partial_{\beta}\bar {g}_{\nu\alpha}) + \frac{1}{2}\bar {g}^{\mu\beta}(\partial_{\nu}{h}_{\alpha\beta} + \partial_{\alpha}{h}_{\nu\beta} - \partial_{\beta}\bar {h}_{\nu\alpha}),\\
    &= \frac{1}{2}\bar {g}^{\mu\beta}(\nabla_{\nu}{h}_{\alpha\beta} + \nabla_{\alpha}{h}_{\nu\beta} - \nabla_{\beta}\bar {h}_{\nu\alpha}),
\end{split}
\end{equation}
where the covariant derivatives $\nabla_{\mu}$ are calculated with respect to the background metric $\bar {g}_{\mu\nu}$.
The first order corrections to the Ricci tensor can be computed from the Palatini identity
\begin{equation}
\begin{split}
    \delta^{(1)}R_{\mu\nu} &= \nabla_{\alpha}(\delta^{(1)}\Gamma^{\alpha}_{\mu\nu}) - \nabla_{\mu}(\delta^{(1)}\Gamma^{\alpha}_{\alpha\nu}),\\
    &= \frac{1}{2}(\nabla_{alpha}\nabla_{\mu}{h}_{\nu}^{\;\alpha} + \nabla_{\alpha}\nabla_{\nu}{h}_{\mu}^{\;\alpha} - \square{h}_{\mu\nu} - \nabla_{\mu}\nabla_{\nu}h)\label{22},
\end{split}
\end{equation}
where $\square := \bar{g}^{\mu\nu}\nabla_{\mu}\nabla_{\nu}$ and $h = \bar{g}^{\mu\nu}h_{\mu\nu}$.
The first order correction to the Ricci scalar is thus computed as
\begin{equation}
\begin{split}
    \delta^{(1)}R &= \delta^{(1)}(g^{\alpha\beta}R_{\alpha\beta}),\\
    &= -h^{\alpha\beta}\bar R_{\alpha\beta} + \bar g^{\alpha\beta}\delta^{(1)}R_{\alpha\beta}\label{23}
\end{split}
\end{equation}
We can use (\ref{22}) to evaluate the last term in (\ref{23}),
\be
\bar{g}^{\alpha\beta}\delta^{(1)}R_{\alpha\beta} = \nabla_{\alpha}\nabla_{\beta}h^{\alpha\beta} - \square{h},
\ee
where we have used the property $\nabla_{\alpha}\bar{g}^{\mu\nu} = 0$.\\
Using these relations we see that Eq. (\ref{20}) can be written as 
\be
-\nabla_{\alpha}\nabla_{\mu}h^{\;\alpha}_{\nu}
-\nabla_{\alpha}\nabla_{\nu}h_{\mu}^{\;\alpha}
+ \square{h}_{\mu\nu}\nabla_{\mu}\nabla_{\nu}h + \bar{g}_{\mu\nu}(\nabla_{\alpha}\nabla_{\beta}h^{\alpha\beta} - \square{h}) + \bar{R}h_{\mu\nu} -\bar{g}_{\mu\nu}h^{\alpha\beta}\bar{R}_{\alpha\beta} = -\frac{\kappa^2}{2}\delta^{(1)}T^{\mu\nu}\label{25}
\ee
We can simplify the equation by making the TT gauge choice $\nabla_{\mu}h^{\mu\nu} = 0$ and $h = 0$.The first term in (\ref{25}) can be written using the commutator identity
\be
\begin{split}
\nabla_{\alpha\mu}\nabla_{\mu}h_{\nu}^{\alpha} &= \nabla_{\mu}\nabla_{\alpha}h_{\nu}^{\alpha} + h_{\nu}^{\alpha}\bar{R}_{\alpha\mu} - \bar{R}_{\alpha\mu\nu\beta}h^{\alpha\beta}\\
&= h_{\nu}^{\alpha}\bar{R}_{\alpha\mu} - \bar{R}_{\alpha\mu\nu\beta}h^{\alpha\beta},
\end{split}
\ee
where the first term in the RHS is zero due to the gauge condition.Similarly the second term in (\ref{25}) can be written as
\be
\nabla_{\alpha\mu}\nabla_{\mu}h_{\nu}^{\alpha} = h_{\mu}^{\alpha}\bar{R}_{\alpha\nu} - \bar{R}_{\alpha\nu\mu\beta}h^{\alpha\beta}\label{27}.
\ee
With these simplifications the wave equation for $h_{\mu\nu}$ TT gauge reduces to
\be
\square{h}_{\mu\nu} + 2\bar{R}_{\alpha\mu\nu\beta}h^{\alpha\beta} + \bar{R}h_{\mu\nu} - \bar{g}_{\mu\nu}h^{\alpha\beta}\bar{R}_{\alpha\beta} - h_{\mu}^{\beta}\bar{R}_{\beta\nu} - h_{\nu}^{\beta}\bar{R}_{\beta\mu} = -\frac{\kappa^2}{2}\delta^{(1)}T^{\mu\nu}\label{28}
\ee
Taking the trace we get the relation
\be
\bar{R}_{\mu\nu}h^{\mu\nu} = \frac{\kappa^2}{8}\bar{g}^{\mu\nu}\delta^{(1)}T_{\mu\nu}\label{29}.
\ee
This can be used to replace the Ricci terms in (\ref{27}) to obtain the wave equation for the tensor perturbations $h_{\mu\nu}$ in the TT gauge,
\be
\square{h}_{\mu\nu} + 
2\bar{R}_{\alpha\mu\nu\beta}h^{\alpha\beta} = \frac{\kappa^2}{4}(h_{\mu}^{\alpha}\bar{T}_{\alpha\nu} + h_{\nu}^{\alpha}\bar{T}_{\alpha\mu} + \frac{1}{2}\bar{g}_{\mu\nu}\bar{g}^{\alpha\beta}\delta^{(1)}T_{\alpha\beta} - 2\delta^{(1)}T_{\mu\nu})\label{30}.
\ee
We will use the master equation (\ref{30}) to calculate the effect of different types of matter on the propagation of gravitational waves.

\subsection{Propagation of Gravitational Waves Through Perfect Fluid Medium}\label{perfect}

We consider gravitational waves in a general background $g_{\mu\nu} = \bar{g}_{\mu\nu} + h_{\mu\nu}$.For cosmological applications $\bar{g}_{\mu\nu}$ is taken to be the flat FLRW metric $\bar{g}_{00} = 1,\bar{g}_{ij} = -{a}^2\delta_{ij}$.The fluid through which the GR waves propagate is taken to have a stress tensor of the perfect fluid form
\be
T_{\alpha\beta} = (\rho + \textit{p})\textit{u}_{\alpha}\textit{u}_{\beta} - \textit{p}{g}_{\alpha\beta}.
\ee
The first order perturbation of $T_{\alpha\beta}$ is given by
\be
\delta^{(1)}T_{\alpha\beta} = -\textit{p}h_{\alpha\beta} + \frac{1}{2}(\rho + \textit{p})\bar{g}_{\alpha\beta}h_{\mu\nu}\textit{u}^{\mu}\textit{u}^{\nu} + (\rho + \textit{p})(\textit{u}_{\alpha}\textit{u}^{\nu}h_{\nu\beta} + \textit{u}_{\beta}\textit{u}^{\nu}h_{\nu\alpha})\label{32}.
\ee
Taking the trace we have the relation
\be
\bar{g}^{\alpha\beta}\delta^{(1)}T_{\alpha\beta} = 4(\rho + \textit{p})h_{\mu\nu}\textit{u}^{\mu}\textit{u}^{\nu}.
\ee
Using (\ref{32}) in the Ricci tensor relation (\ref{29}) we obtain
\be
\bar{R}_{\alpha\beta}h^{\alpha\beta} = \frac{\kappa^2}{4}(\rho + \textit{p})h_{\mu\nu}\textit{u}^{\mu}\textit{u}^{\nu}
\ee
From these two equations, we have the consistency relations
\be
(\rho + \textit{p})h_{\mu\nu}\textit{u}^{\mu}\textit{u}^{\nu} = 0.
\ee
Using these relations in (\ref{30}) we obtain the wave equation of graviton propagating through a perfect fluid medium
\be
\square{h}_{\alpha\beta} + 2\bar{R}_{\mu\alpha\beta\nu}h^{\mu\nu} + \frac{\kappa^2}{4}(\rho + \textit{p})(\textit{u}_{\alpha}\textit{u}^{\nu}h_{\nu\beta} + \textit{u}_{\beta}\textit{u}^{\nu}h_{\nu\alpha}) = 0,
\ee
where the D’Alembertian $\square$ and the Riemann tensor $\bar{R}_{\mu\alpha\beta\nu}$ are defined w.r.t the background metric $\bar{g}_{\mu\nu}$.In the case of a comoving fluid $\textit{u}^{0} = 1,\textit{u}^{i} = 0$ and for the TT gauge $h_{\mu0} = 0$ the terms $\textit{u}^{\nu}h_{\nu\beta} = 0$ and the wave equation reduces to
\be
\square{h}_{\alpha\beta} + 2\bar{R}_{\mu\alpha\beta\nu}h^{\mu\nu} = 0.
\ee
For the FRLW metric $\bar{g}_{00} = 1,\bar{g}_{ij} = -{a}^2\delta_{ij}$ we have
\be
\bar{\Gamma}^{i}_{0j} = \textit{H}\delta^{i}_{j},   \bar{\Gamma}^{0}_{ij} = a^2\textit{H}\delta_{ij},   \bar{R}_{kijl}h^{kl} = \textit{H}^2h_{ij},
\ee
where $\textit{H} = \dot{a}/a$, and the gravitational wave equation in the FLRW universe is 
\be
\Ddot{h}^{i}_{j} + 3\textit{H}\dot{h}^{i}_{j} - \frac{\partial^2}{a^2}h^{i}_{j} = 0,
\ee
where $h^{i}_{j} = \bar{g}^{ik}h_{kj}$ and where $\partial^2 = \delta^{ij}\partial_{i}\partial_{j}$.The speed of gravitational waves $c_g$ is the coefficient of the $\frac{\partial^2}{a^2}$ term and in this case $c_g = 1$.We see that despite the non-zero energy density and pressure of the fluid medium gravitational waves propagate through the medium with the speed of light.

\section*{Acknowledgement} Support by the TEBIP High Performers Program of
the Board of Higher Education of Turkey is gratefully acknowledged.\\

\section*{Remarks}

\textbf{Contribution Statement:} All authors contributed equally to the whole work including calculations, analysis and manuscript typing.

\textbf{Data Availability Statement:} Data supporting this study are included within the article and/or supporting materials.

\textbf{Competing Interests:} The authors declare no competing interests.

\bibliography{references}

\end{document}